\documentclass[12pt]{article}

\usepackage{amssymb}
\usepackage{amsmath}
\usepackage{times}
\usepackage{graphicx}
\usepackage{float}
\usepackage{xcolor}
\usepackage{subfigure}
\usepackage{cite}

\usepackage{hyperref}
\numberwithin{equation}{section}
\hypersetup{
  colorlinks=true,
  linkcolor=blue,
  citecolor=blue,
  urlcolor=blue,
  filecolor=blue,
  linktoc=page,
  pdfborder={0 0 0},
  pdftitle={Dual gravities from entanglement entropy},
  pdfauthor={Jaehyeok Huh and Chanyong Park}
}

\begin{document}

\baselineskip 6mm
\renewcommand{\thefootnote}{\fnsymbol{footnote}}

%------ Chanyong Park's macros, etc  ------

\newcommand{\nc}{\newcommand}
\newcommand{\rnc}{\renewcommand}

%%%%%%%%%%%%%%%%%%%%%%%%%%%% 
%                                                                                     %
%                NEW COMMANDS AND MACROS               %
%                                                                                     %
%%%%%%%%%%%%%%%%%%%%%%%%%%%% 

\newcommand{\tcb}[1]{#1}
\newcommand{\tcr}[1]{#1}
\newcommand{\tcg}{\textcolor{green}}
\newcommand{\app}[1]{Appendix~\ref{#1}}

%%%%% Simplify some frequently used LaTeX commands %%%%%

\def\be{\begin{eqnarray}}
\def\ee{\end{eqnarray}}
\def\nn{\nonumber\\}

%%%%%  Temporary notation %%%%

\def\ct{\cite}
\def\la{\label}
\def\eq#1{\textcolor{blue}{(\ref{#1})}}
\def\app#1{Appendix \ref{#1}}

%%% Greek letters %%%

\def\a{\alpha}
\def\b{\beta}
\def\g{\gamma}
\def\G{\Gamma}
\def\d{\delta}
\def\D{\Delta}
\def\e{\epsilon}
\def\et{\eta}
\def\ph{\phi}
\def\Ph{\Phi}
\def\ps{\psi}
\def\Ps{\Psi}
\def\k{\kappa}
\def\l{\lambda}
\def\L{\Lambda}
\def\m{\mu}
\def\n{\nu}
\def\th{\theta}
\def\Th{\Theta}
\def\r{\rho}
\def\s{\sigma}
\def\S{\Sigma}
\def\ta{\tau}
\def\o{\omega}
\def\O{\Omega}
\def\pr{\prime}

%%%%% Mathematical Symbols

\def\half{\frac{1}{2}}

\def\goto{\rightarrow}

\def\na{\nabla}
\def\grad{\nabla}
\def\curl{\nabla\times}
\def\div{\nabla\cdot}
\def\pa{\partial}
\def\fr{\frac}

\def\bra{\left\langle}
\def\ket{\right\rangle}
\def\lb{\left[}
\def\lc{\left\{}
\def\ls{\left(}
\def\lp{\left.}
\def\rp{\right.}
\def\rb{\right]}
\def\rc{\right\}}
\def\rs{\right)}

\def\vac#1{\mid #1 \rangle}

%%%%  Special symbol

\def\td#1{\tilde{#1}}
\def\check{ \maltese {\bf Check!}}

%%%%% Roman point in math

\def\Tr{{\rm Tr}\,}
\def\det{{\rm det}}
\def\inf{\infty}

%%%%% Special format

\def\bc#1{\nnindent {\bf $\bullet$ #1} \\ }
\def\ch {$<Check!>$ }
\def\ss {\vspace{1.5cm}}
\def\inf{\infty}

\begin{titlepage}

%-------- preprint number --------
\hfill\parbox{2cm} { }

\hspace{13cm} %\today
 
\vspace{2cm}

\begin{center}
%------------ title ------------
{\Large \bf Dual gravities from entanglement entropy}

%-------- authors and addresses --------
\vskip 1. cm
{Jaehyeok Huh $^{a}$\footnote{e-mail gj$\_$wogur@gm.gist.ac.kr} and Chanyong Park$^{a}$\footnote{e-mail cyong21@gist.ac.kr}}

\vskip 0.5cm

{\it $^a\,$ Department of Physics and Photon Science, Gwangju Institute of Science and Technology,  123 Cheomdangwagi-ro, Buk-gu, Gwangju  61005, Korea  } \\

\end{center}

\thispagestyle{empty}

\vskip1cm

%------------ abstract -----------

\centerline{\bf ABSTRACT} \vskip 4mm

We study how entanglement entropy can be used to reconstruct a dual gravity description. For a two-dimensional thermal system, equal-time entanglement entropy determines the spatial part of a three-dimensional black-hole geometry within a given metric ansatz. For an undeformed conformal field theory, the same inverse problem can be solved analytically by the Abel transformation. For a relevant deformation, we work in three-dimensional Einstein gravity with one canonical scalar field whose profile is monotonic. The geometry reconstructed from the given entanglement entropy then allows us to determine the scalar profile and the part of the scalar potential sampled by the RG flow. It also determines the holographic $\beta$-function and $c$-function.

%PACS numbers:

%\today

\end{titlepage}

\renewcommand{\thefootnote}{\arabic{footnote}}
\setcounter{footnote}{0}

\begingroup
\hypersetup{linkcolor=blue,citecolor=blue}
\tableofcontents
\endgroup

%%%%%%%%%%%%%%%%%%%%%%%%%%%% 
%                                                                                    %
%                              Sec.  Introduction                        %
%                                                                                     %
%%%%%%%%%%%%%%%%%%%1%%%%%%%%% 

\section{Introduction}
\label{sec:intro}

The Anti-de Sitter (AdS)/conformal field theory (CFT) correspondence relates a gravitational theory in an asymptotic AdS space to a quantum field theory (QFT) defined on its boundary \cite{Maldacena:1997re, Gubser:1998bc, Witten:1998qj, Witten:1998zw}. This relation gives a useful geometric description of strongly coupled quantum systems. An important example is the Ryu-Takayanagi formula \cite{Ryu:2006bv, Ryu:2006ef}
\be
S_E = \fr{{\rm Area}(\g_A)}{4 G_N} .
\ee
It maps the entanglement entropy of a boundary subsystem to the area of a minimal surface in the bulk. Standard field-theory results and general properties of entanglement entropy provide the corresponding boundary description \cite{Calabrese:2004eu, Calabrese:2009qy, Headrick:2007km, Casini:2011kv}. The subsystem size then plays the role of a length scale in the dual QFT.

The holographic radial direction is also related to the renormalization group scale \cite{Henningson:1998ey, Freedman:1998tz, Gubser:1999vj, deBoer:1999tgo, Verlinde:1999xm, deBoer:2000cz, Akhmedov:1998vf, deHaro:2000vlm, Papadimitriou:2004ap, Papadimitriou:2010as, Heemskerk:2010hk, Skenderis:2002wp}. This relation has been used to study RG flows, entanglement entropy, and the emergence of bulk geometry from boundary data \cite{Girardello:1998pd, Ryu:2006bv, Ryu:2006ef, Lewkowycz:2013nqa, Hubeny:2007xt, Nishioka:2009un, Nozaki:2012zj, Rangamani:2016dms, Faulkner:2013ana, Swingle:2009bg, Park:2015afa, Kim:2016hig, Kim:2017lyx, Narayanan:2018ilr, Park:2018snf, Park:2019pzo, Park:2022mxj, Park:2025ahy}. The relation between entanglement and bulk gravitational dynamics has also been developed from complementary viewpoints \cite{VanRaamsdonk:2009ar, VanRaamsdonk:2010pw, Lashkari:2013koa, Faulkner:2013ica, Faulkner:2017tkh}.

The inverse problem asks whether the bulk geometry can be obtained from boundary observables. Kinematic-space and integral-geometric methods reconstruct geometric information from boundary geodesics and entanglement data \cite{Czech:2015qta, Balasubramanian:2013lsa, Balasubramanian:2013rqa, Huang:2015bza, Bhattacharya:2025hag, Espindola:2018ozt, Espindola:2017jil}. Closely related metric-reconstruction approaches use extremal-area variations, noisy entanglement data, lattice entanglement data, and combinations of entanglement with other observables \cite{Bao:2019bib, Jokela:2020auu, Hashimoto:2021umd, Jokela:2023rba, Xu:2023eof, Jokela:2025ime, Ji:2025vks}. Numerical and machine-learning methods provide another route when analytic inversion is difficult \cite{Hashimoto:2018ftp, Akutagawa:2020yeo, Park:2022fqy, Park:2023slm, Ahn:2024gjf, Ahn:2024jkk, Ahn:2025tjp}. The authors of Refs.~\cite{Hammersley:2007ab, Hammersley:2006cp, Bilson:2008ab, Bilson:2010ff} studied closely related inverse constructions of bulk metric functions. The Abel inversion used in this work follows these previous studies and has the same inverse structure. In the present work, however, after reconstructing the metric factor, we use the Einstein-scalar equations to determine the scalar profile and the part of the potential traced by the flow. The authors of Ref.~\cite{Lu:2025jgk} used pole-skipping of boundary Green functions to reconstruct near-horizon geometric data. Their observable and reconstruction problem are different from the entanglement-based method considered here.

In this work, the phrase dual-gravity reconstruction has a restricted meaning. We assume three-dimensional two-derivative Einstein gravity with one canonical scalar field together with the metric ansatz and boundary conditions stated below. Entanglement entropy does not determine an arbitrary gravitational action. Within this model class, it determines the spatial metric and, for a monotonic RG flow, allows us to infer the scalar profile and the sampled part of the scalar potential. We first study a finite-temperature example. Equal-time entanglement entropy reconstructs the spatial radial geometry. The temporal metric component and the Hawking temperature require an additional bulk completion. We then consider a zero-temperature relevant deformation. The Abel transformation gives the warp factor, which is then used in the Einstein-scalar equations to determine the scalar sector and RG observables such as the $\beta$- and $c$-functions.

The paper is organized as follows. Sec.~\textcolor{blue}{2} discusses the thermal system and the reconstruction of its spatial black-hole geometry. Sec.~\textcolor{blue}{3} studies the scalar deformation and reconstructs the corresponding RG flow. Sec.~\textcolor{blue}{4} gives the discussion.

\section{Holographic dual geometries of thermal systems}

We consider a $(d+1)$-dimensional AdS black hole with $N$ multiple hairs, which is governed by the following action
\be
S = \fr{1}{2 \k^2} \int d^{d+1} x \sqrt{-G}  \  \ls {\cal R} - 2 \L \rs  + S_{matter}  ,
\ee
where the conserved charges derived from a matter action $S_{matter}$ provide $(N-1)$ hairs. Then, the Einstein equation is reduced to
\be
{\cal R}_{MN} - \half {\cal R} \, G_{MN} + \L \, G_{MN} = T_{M N}  ,
\ee
where $T_{MN}$ is the energy-momentum tensor derived from $S_{matter}$. To describe a black hole, which corresponds to a thermal system on the dual field side, we use the following metric ansatz
\be
ds^2 = \fr{R^2}{z^2}  \ls - f(z) dt^2 + \fr{1}{f(z)}  dz^2  + \d_{ij} dx^i dx^j  \rs ,
\ee 
where $f(z)$ approaches one at the boundary ($z \to 0$), so that the geometry is asymptotically AdS. At finite temperature we additionally assume the Schwarzschild gauge relation $g_{tt}g_{zz}=-R^4/z^4$, so that the same blackening function $f(z)$ appears in $g_{tt}$ and $g_{zz}$. Equal-time entanglement entropy directly constrains the spatial component $g_{zz}$, but does not independently determine $g_{tt}$. Consequently, the Hawking temperature follows only after imposing this metric ansatz. It is not inferred from equal-time entanglement data alone.  If there is no $S_{matter}$, an AdS Schwarzschild black hole appears as a general solution with
\be
f(z)  = 1 - M \, z^d ,
\ee
where $M$ indicates a black hole mass. On the dual QFT side, the black hole mass is reinterpreted as the energy of massless radiation.

Adding more matter $S_{matter}$ further generalizes the Schwarzschild black hole into a multi-hair black hole, for example, a $p$-brane gas geometry \cite{Park:2020jio, Park:2021wep, Park:2022abi}. The energy-momentum tensor of $(p+1)$-branes is proportional to the number density of $(p+1)$-branes, which deforms the blackening factor to be
\be
f(z)  = 1 - \sum_{p=0}^{d-2} \r_p \, z^{d-p-1}- M \, z^d    ,   \la{Result: BHmetricF}
\ee
where we omit the case with $p=d-1$ because $d$-branes shift the background cosmological constant. On the dual QFT side, $\r_p$ is reinterpreted as the number density of $p$-dimensional solitonic objects. For example, $\r_0$ indicates the number density of heavy particles, whereas $\r_1$ is the number density of solitonic $1$-dimensional objects on the dual QFT side.

 In holography, the black hole geometry encodes the thermodynamics of the dual thermal state. The Hawking temperature and the Bekenstein-Hawking entropy follow directly from the metric
\be
T_H &=& -  \fr{f'(z_h)}{4 \pi}  \quad {\rm and} \quad S_{BH} = \fr{V}{4 G} \fr{R^{d-1}}{z_h^{d-1}} ,  \la{Result: blackhth}
\ee 
where $V$ indicates a $(d-1)$-dimensional regularized volume. From the holography point of view, the black hole's temperature and entropy are reinterpreted as those of the dual QFT. The other thermodynamic quantities can be derived by applying the thermodynamic law \cite{Park:2022abi}
\be
d U  (S_{BH},N_a,V)  = T_H  \,  d S_{BH} + \sum_a \m_a \, d N_a -P \, dV  ,  \la{Relation: Thermo}
\ee
where $N_a$ are $n$ conserved quantities. We include the particle number $N_0=\r_0 V$ among them. More generally, $N_p=\r_p V$ for $p=0,\ldots,n-1$. Thus derivatives with respect to $V$ below are taken at fixed $N_a$, in particular at fixed $N_0$, rather than at fixed density $\r_0$. As a result, when the volume is fixed, the thermal system is characterized by temperature and matter densities, and then other thermodynamic quantities are given as functions of thermodynamic variables, $T_H$ and $\r_p$.  

Given $N_a$ and $V$, the internal energy of the thermal system is derived by
\be
U (S_{BH},N_a,V) = \int T_H  \, dS_{BH} = \fr{(d-1) \, R^{d-1} V}{16 \pi G} \int d z_h  \fr{f'(z_h)}{z_h^d}  .
\ee
After the Legendre transformation, the free energy also reads
\be
F (T_H,N_a,V)  = U - T_H \, S_{BH}  .
\ee
Define the ground state energy and pressure at zero temperature
\be
U_0 &=& \lim_{T_H \to 0} U , \nn
P_0 &=& \lim_{T_H \to 0} P = -  \lim_{T_H \to 0} \lp \fr{\pa F}{\pa V} \right|_{T_H, N_a} .
\ee
Then, the thermal energy and thermal pressure become
\be
U_{th} = U - U_0 \quad {\rm and} \quad P_{th} = P - P_0 .
\ee
From these thermodynamic quantities, we evaluate some physical quantities characterizing the thermal system, like an equation of state and a specific heat
\be
w = \fr{P_{th} V}{U_{th}}   \quad {\rm and} \quad c_V = \fr{\pa U_{th}}{ \pa T_H}  .
\ee

Now, we consider the entanglement entropy of the dual thermal system. To evaluate the entanglement entropy, we consider a strip-shaped subsystem, $-\ell/2 \le x^1 = x \le \ell/2$ and $- L/2 \le x^i \le L/2$ with $L \to \infty$. In the holographic setup, its entanglement entropy is described by a minimal surface extending to the dual geometry. This is called the Ryu-Takayanagi formula. Then, the entanglement entropy is determined by the area of a minimal surface \cite{Ryu:2006bv, Ryu:2006ef, Hubeny:2007xt}
\be
S_E = \fr{R^{d-1} \, V_{d-2}}{4 G} \int_{-\ell/2}^{\ell/2} dx  \fr{ \sqrt{z'^2 + f} }{z^{d-1} \sqrt{f} } ,
\ee
where $V_{d-2}= L^{d-2}$ and the prime means a derivative with respect to $x$. Using conserved quantities and introducing a turning point $z_t$ where $z' = 0$, we determine the subsystem size and the entanglement entropy as functions of the turning point
\be
\ell (z_t) &=& \int_{\e}^{z_t} dz \fr{2 z^{d-1}}{\sqrt{f(z)} \ \sqrt{z_t^{2 (d-1)} - z^{2(d-1)} } } , \nn
S_E (z_t)&=& \fr{R^{d-1} \, V_{d-2}}{2 G} \int_{\e}^{z_t} dz \ \fr{z_t^{d-1}}{z^{d-1} \sqrt{f(z)} \ \sqrt{z_t^{2(d-1)} - z^{2 (d-1)}} } ,   \la{Relation: EEell}
\ee
where $\e$ is introduced as a UV cutoff. Here, we can remove the UV cutoff by applying an appropriate renormalization procedure. In this work, for convenience, we introduced a small physical UV cutoff $\e=10^{-3}$, which is reinterpreted as the lattice spacing in condensed matter theory.

Combining the above two relations and removing the turning point, we can also determine the entanglement entropy as a function of the subsystem size, $S_E (\ell)$. In this case, the entanglement entropy always satisfies the following relation \cite{Park:2022fqy, Park:2023slm}
\be
\fr{d S_E}{ d \ell} = \fr{ V_{d-2}}{4 G}  \fr{R^{d-1} }{z_t^{d-1}}  .
\ee
In an infinite subsystem size limit ($\ell \to \infty$ and $z_t \to z_h$), the entanglement entropy is reduced to the Bekenstein-Hawking  entropy
\be
S_{BH} = \lim_{z_t \to z_h} S_E =  \fr{ V}{4 G}  \fr{R^{d-1} }{z_h^{d-1}}   ,   \la{Relation: BHentr}
\ee
where a \tcr{$(d-1)$}-dimensional volume is given by $V = \ell L^{d-2}$.

\subsection{Multiple-hairs black hole as the effective theory of a dual thermal system}

We now solve the inverse problem. The entanglement entropy is taken as input, and the metric function is obtained from the holographic integral equation in Eq.~\eq{Relation: EEell}. Related inverse problems have also been studied with machine-learning methods \cite{Park:2022fqy, Park:2023slm, Kim:2025aln}. We assume that the QFT has a UV fixed point. The corresponding bulk geometry is asymptotically AdS and satisfies $f(0)=1$.  

Now, we parametrize the radial coordinate as
\be
z_i = \e + i \, \d z  \la{Relation:radialco}
\ee
with
\be
\d z = \fr{z_{h} - \e}{N}
\ee
where $N$ is a large integer. Then, the range of $z$ is restricted to $\e \le z \le z_h$, where $\e$ is interpreted as the lattice spacing. According to the holography principle in Eq.~\eq{Relation: EEell}, the subsystem size and entanglement entropy are determined by the radial coordinate in Eq.~\eq{Relation:radialco}
\be
\ell_n &=& \sum_{i=1}^{n} \int_{z_{i-1}}^{z_i} dz \fr{2 z}{\sqrt{ \bar{f} (z_i)} \ \sqrt{z_n^2 - z^2} }   , \nn
S_n &=& \sum_{i=1}^{n} \fr{R}{2 G}   \int_{z_{i-1}}^{z_i} dz \ \fr{z_n}{z \sqrt{ \bar{f}  (z_i)} \ \sqrt{z_n^2 - z^2} }  ,   \la{Relation:EEell}
\ee
where  $z_n$ parametrises the turning point and $\bar{f} (z_i)$ is defined as the average value of $f(z_{i-1})$ and $f(z_i) $
\be
\bar{f}  (z_i) = \fr{f(z_{i-1}) + f(z_i) }{2} .
\ee

After imposing the boundary condition, $f(z_0) = 1$ where $z_0 = \e$, we can find $f(z_i)$ successively.  For example, when we set $n=1$, $\ell_1$ and $S_1$ depend only on $f(z_1)$. Combining these results, we find $f(z_1)$ satisfying $\ell_1 = \ell (z_1)$ and $S_1 = S_E (z_1)$ simultaneously. Using the known $f(z_0)$ and $f(z_1)$, we also find $f(z_2)$ satisfying $\ell_2 = \ell (z_2)$ and $S_2 = S_E (z_2)$. Repeating this procedure, we finally obtain $f(z_i)$ for $i=1, \cdots, n$. Then, we derive the metric factor from the entanglement entropy data $S_E (\ell)$
\be
\bar{f}  (\bar{z}_i) = \fr{ f(z_{i-1}) + f(z_i) }{2} \quad {\rm at} \quad \bar{z}_i = \fr{  z_{i-1} + z_i }{2} ,
\ee
which corresponds to the average value of $f(z)$ in the range of $z_{i-1} \le z \le z_i$.

%%%%%%%%%%%%%%%%%%%%%%%%
\begin{figure}%[h!]
\begin{center}
\vspace{-0.5cm}
\hspace{-0.5cm}
\subfigure[Entanglement entropy (input)]{ \includegraphics[angle=0,width=0.45 \textwidth]{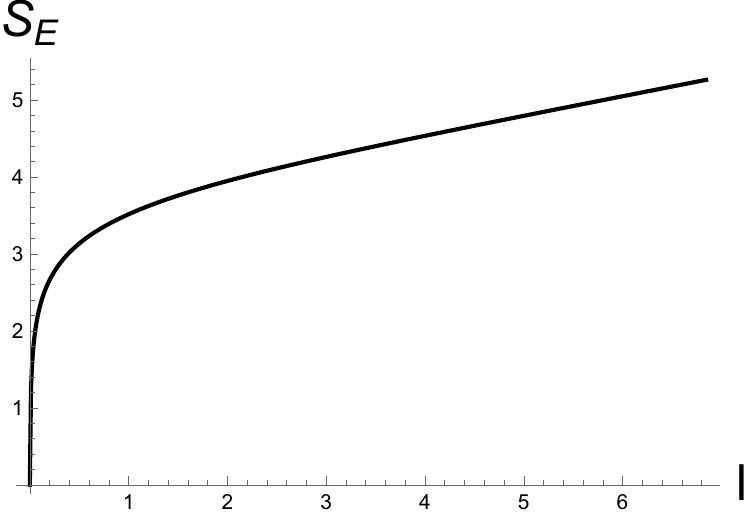}}
\hspace{0.5cm}
\subfigure[Blackening factor (output)]{ \includegraphics[angle=0,width=0.45 \textwidth]{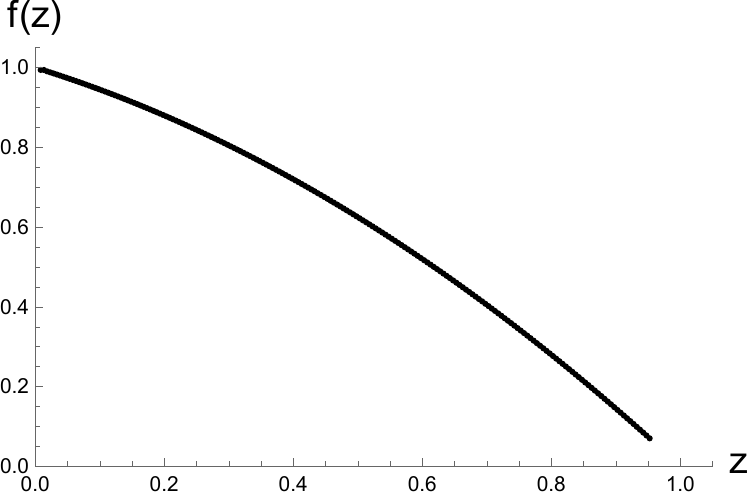}}
\vspace{-0cm}
\caption{ (a) The entanglement entropy relying on the subsystem size. We evaluate the entanglement entropy holographically for $M=\r_0 =1/2$ with $R=G=z_h=1$, $\e= 10^{-3}$, and $N=20$. (b) When the entanglement entropy is given without knowing the values of $M$ and $\r_0$, we reconstruct the dual black hole geometry from the entanglement entropy as the input data.  }
%\label{number}
\end{center}
\end{figure}
%%%%%%%%%%%%%%%%%%%%%%%%%

\subsection{Thermodynamic quantities from reconstructed black hole geometry}

Let us first consider the benchmark entanglement entropy $S_E(\ell)$ shown in Fig.~\textcolor{blue}{1} (a). Its large-$\ell$ volume law is characteristic of a thermal state. This behavior does not uniquely determine the microscopic matter content. Therefore, we assume a model consisting of radiation and massive particles to generate the input data. 

Following the previous holographic description, we numerically reconstruct the blackening factor in Fig.~\textcolor{blue}{1} (b), which reproduces the given entanglement entropy data in Fig.~\textcolor{blue}{1} (a) up to small numerical errors. Using the extrapolation of the numerical blackening factor in Fig.~\textcolor{blue}{1} (b), we find that the event horizon appears at
\be
z_h  \approx 0.99986  .   \la{Result: Nzh}
\ee  
After specifying the temporal completion of the metric, we evaluate the slope of $f(z)$ at the event horizon and determine the Hawking temperature  
\be
T_H =  -   \fr{ f' (z_h)  }{4 \pi} \approx  0.12054 \la{Result: NTH}
\ee
From these numerical results, we can further determine the other parameter $\r_0$. Within the ansatz specified above, the dual black hole, according to the holographic relation in Eq.~\eq{Result: BHmetricF}, has the following form
\be
f (z) =  1 - \r_0 \, z - M  \, z^2 ,
\ee
where the values of $\r_0$ and $M$ are not yet determined. Since the black hole must have an event horizon satisfying $f(z_h) = 0$,  the black hole mass is rewritten as a function of $z_h$ and $\r_0$. Therefore, the blackening factor is reexpressed as
\be
f (z) =  1 - \r_0  \, z - \fr{1 - \r_0  \, z_h}{z_h^2}  \,  z^2  .
\ee
Using this metric form together with the numerical values, $z_h$ in Eq.~\eq{Result: Nzh} and $T_H$ in Eq.~\eq{Result: NTH}, the density of massive particles and the black hole mass are determined as
\be
\r_0 &=& \fr{2}{z_h} - 4 \pi T_H \approx  0.48553 , \nn
M &=& \fr{1 - \r_0  \, z_h}{z_h^2}  \approx 0.51468   .
\ee
As a result, the obtained numerical values determine the blackening factor, which is the dual geometry of the given entanglement entropy data
\be
f (z) =  1 - 0.48553  \, z - 0.51468  \, z^2 .   \la{Result: ReCMetric}
\ee
Within the assumed black hole model, the reconstructed spatial geometry can then be used to evaluate thermodynamic quantities \cite{Bardeen:1973gs, Bekenstein:1973ur, Hawking:1975vcx, Wald:1993nt, Carlip:1995qv, Freedman:1999gp}, although these quantities are not determined by equal-time entanglement entropy alone. Applying the thermodynamic law in Eq.~\eq{Relation: Thermo} to the reconstructed black hole geometry in Eq.~\eq{Result: ReCMetric}, we obtain the internal energy associated with the entanglement data in Fig.~\textcolor{blue}{1} (a)
\be
U =  \fr{ 16 \pi^2  T_H^2 }{64 \pi }  \,  \fr{R  V}{G}  - \fr{\r_0^2    }{64 \pi }    \,  \fr{R  V}{G} .
\ee
Here, the first term corresponds to the thermal internal energy 
\be
U_{th} = \fr{\pi T_H^2  }{4 }    \,  \fr{R  V}{G} \approx 0.01141 \,  \fr{R  V}{G} ,
\ee
while the last term describes the ground state energy independent of temperature. From $F=U-T_HS_{BH}$ we obtain
\be
F=-\fr{R V}{64\pi G}\left(4\pi T_H+\r_0\right)^2
=-\fr{R}{64\pi G}\left(16\pi^2T_H^2V+8\pi T_HN_0+\fr{N_0^2}{V}\right).
\ee
Therefore, the pressure of this system is reduced to
\be
P=-\left.\fr{\pa F}{\pa V}\right|_{T_H,N_0}
=\fr{R}{64\pi G}\left(16\pi^2T_H^2-\r_0^2\right)
\quad {\rm and} \quad
P_0=-\fr{R\r_0^2}{64\pi G},
\ee
where $P_0$ denotes the vacuum pressure. Here, $N_0=\rho_0 V$ denotes the conserved particle number, so $\rho_0=N_0/V$ varies under the volume derivative. Then, the thermal pressure can be defined as
\be
P_{th}=P-P_0=\fr{\pi T_H^2}{4}\fr{R}{G}
\approx0.01141\fr{R}{G}.
\ee
The resulting equation of state and specific heat are
\be
w &=& \fr{P_{th}V}{U_{th}}=1, \nn
c_V &=& \fr{\pi T_H}{2}\fr{RV}{G}\approx0.18934\fr{RV}{G}>0.
\ee
Here, the specific heat is positive, so the thermal system we consider is thermodynamically stable \tcr{when the particle number and volume are fixed}. In Table \textcolor{blue}{1}, we summarize the obtained numerical results and compare them with the true values.

%%%%%%%%%%%%%%%%%%%%%%%%%%%%
\vspace{0.5cm}
\begin{center}
	\begin{tabular}{|c||c|c|c|}
		\hline
&  Derived value &  True value & Error  \\
\hline \hline
$U_{th}$	& $0.01141 $ &  $0.01119$ &  1.97 \% \\
\hline
$P_{th}$	& $0.01141$  &  $0.01119$  &  1.97 \% \\
\hline
$w$	& $1.00000$  &  $1.00000$ &  0.00 \% \\
\hline
$c_V$	    & $0.18934 $ &  $0.18750$ & 0.98 \%  \\
\hline 
	\end{tabular}
\end{center}
\vspace{0.5cm}
\noindent {\bf Table 1}. \  We summarize the thermodynamic quantities derived from the reconstructed dual black hole geometry with $R=V =G=1$ and compare them to the true values derived with $M=\r_0=1/2$.
%%%%%%%%%%%%%%%%%%%%%%%%%%%%

\section{Holographic dual of deformed CFTs by a scalar operator}

We next ask how much of the bulk theory can be inferred after the geometry is reconstructed. We consider a zero-temperature two-dimensional CFT deformed by a scalar operator \cite{Park:2021nyc, Park:2023xmd}. Related geometric reconstruction has also been studied with data-driven methods \cite{Hashimoto:2018ftp, Akutagawa:2020yeo, Park:2022fqy, Park:2023slm, Ahn:2024gjf, Ahn:2024jkk, Seo:2023kyp, Kim:2024car, Kim:2025aln}
\be
{\cal S}_{QFT} = {\cal S}_{CFT} [\Ph] + \int d^2 x \ g \, O[\Ph]   ,
\ee
where $g$ and $O[\Ph]$ indicate a coupling constant and deformation operator, respectively. \tcr{Near the undeformed fixed point, a relevant deformation with $\D<2$ grows as the RG scale is lowered. It drives the theory away from the UV fixed point and can lead to different IR physics. For an irrelevant deformation with $\D>2$, a small dimensionless coupling decreases as the RG scale is lowered. The fixed point is therefore attractive in this direction toward the IR. We define the dimensionless coupling $\hat g=\m^{\D-d}g$, where $d=2$ and $\m$ is the RG scale. The undeformed fixed point corresponds to $\hat g=0$. For the deformation, we consider the branch $\hat g>0$. Near the fixed point, the $\beta$-function is}
\be
\tcr{\b = \fr{\pa \hat g}{\pa \log \m} = - \ls d - \D \rs \hat g + \cdots} ,
\ee
\tcr{The linear term describes the leading behavior near the fixed point, and the ellipsis denotes terms of higher order in $\hat g$.}

This deformed CFT can be understood as holographic dual gravity. According to the holographic principle, a two-dimensional CFT is mapped to a three-dimensional AdS space. In this case, a scalar operator of CFT is mapped to a bulk scalar field in the holographic setup, which also deforms the AdS background geometry. Therefore, the holographic dual of the deformed CFT can be described by the following Einstein-scalar gravity 
\be
{\cal S}_{gr} = \fr{1}{2 \k^2} \int d^3 x \sqrt{-G}  \ls {\cal R} - 2 \L_3 - \fr{1}{2}  \pa_M \ph \  \pa^M \ph  - \fr{V(\ph)}{R^2}  \rs ,   \la{Ansatz: dual gravity}
\ee
where $\L_3=-1/R^2$ and $V(\ph)$ is dimensionless. We restrict the reconstruction to three-dimensional two-derivative Einstein gravity with one canonical scalar. This model is taken as an input since other bulk actions may support the same geometry. Using the standard holographic dictionary, we identify the boundary scalar mode with the coupling constant in the dual theory. Then, the bulk scalar profile can be mapped to the RG flow of the dual theory. \tcr{For an irrelevant operator with $\D>2$ in standard quantization, the corresponding bulk scalar has positive mass squared. The scalar potential therefore has a local minimum at $\ph=0$, satisfying}
\be
\tcr{\fr{\pa^2 V(\ph)}{\pa \ph^2}\bigg|_{\ph=0} > 0}.
\ee
\tcr{The constant solution $\ph=0$ describes the undeformed CFT and has no nontrivial RG flow. A relevant deformation, however, corresponds to negative scalar mass squared and a local maximum of the potential at $\ph=0$, satisfying}
\be
\tcr{\fr{\pa^2 V(\ph)}{\pa \ph^2}\bigg|_{\ph=0} < 0}.
\ee
For a relevant deformation, the scalar field moves toward another minimum of the potential. Its radial evolution represents the RG flow of the dual QFT, and its backreaction changes the bulk geometry. We use this relation to reconstruct the scalar sector from the QFT entanglement data.

\subsection{Undeformed CFT} 

We first consider the undeformed UV CFT. Since there is no nontrivial RG flow, its dual geometry should be pure AdS. In this case the entanglement entropy is known analytically, so the inverse problem can also be solved analytically. 

At zero temperature we assume that the scalar deformation preserves boundary Lorentz symmetry. We also assume a monotonic scalar profile, so that $r(\phi)$ is single-valued. A convenient domain-wall metric is 
\be		\la{met:normalcoord}
ds^2 = dr^2 + \m(r)^2 \ls - dt^2 + d x^2 \rs  ,   \la{Metric: exact}
\ee
where $\m(r) = e^{A(r)}$ can be reinterpreted as the RG scale of the dual QFT. Here, the radial coordinate $r$ extends from $-\infty$ at IR to $\infty$ at UV. Thanks to the AdS/CFT correspondence, the entanglement entropy of the dual QFT is determined by the area of a minimal surface \cite{Ryu:2006bv, Ryu:2006ef}. From the above general metric ansatz in Eq.~\eq{Metric: exact}, the entanglement entropy is governed by  
\be
S_E = \fr{1}{4 G} \int_{- \ell/2}^{\ell/2} dx \sqrt{ r'^2 + e^{2 A(r) }} ,
\ee
where the prime denotes a derivative with respect to $x$. Using the conserved quantity, the subsystem size $\ell$ and entanglement entropy are determined as functions of a turning point $r_t$, at which $r'=0$,
\be
\ell (r_t) &=& \int_{r_t}^{r_\L} dr \ \fr{2 \,  e^{ A (r_t)} }{ e^{A (r)} \sqrt{e^{2 A  (r)} -   e^{2 A (r_t)}}} , 
\la{Relation:subsize} \\
S_E (r_t ) &=& \fr{1}{2 G}\int_{r_t}^{r_\L} d r \ \fr{ e^{A (r)}  }{  \sqrt{ e^{2 A (r)} - e^{2 A (r_t)} }}  ,
\ee
where $r_\L$ denotes an appropriate UV cutoff. When entanglement entropy is given, solving these two integral equations can fix $A(r)$, as done in the previous section. We next use the Abel transformation to extract the dual geometry more analytically from the entanglement entropy.

The above entanglement entropy and the subsystem size are reexpressed as the integral forms of the radial coordinate
\be
S_E (r_t) &=& \fr{1}{4 G} \int_{- \ell/2}^{\ell/2} dx \sqrt{ r'^2 + e^{2 A(r) }}  = \fr{1}{2 G} \int_{r_t}^{r_\L}  dr \sqrt{1+ e^{2 A(r) }  \ls \fr{d x}{dr} \rs^2}  ,  \nn
\ell (r_t) &=& \int_{- \ell/2}^{\ell/2} dx = 2 \int_{r_t}^{r_\L} dr  \ \fr{d x}{dr}  .
\ee 
Since $dx/dr = 1/r' = \infty$ at the turning point $r=r_t$, we find the following relation
\be
\fr{d S_E}{d \ell} = \fr{d S_E/ d r_t}{d \ell / d r_t}  = \fr{\m_t}{4 G}   ,   \la{Relation: ellmu}
\ee 
where $\m_t = e^{A(r_t)}$. This provides us with information about the relation between the metric factor and the subsystem size, $\m_t (\ell)$ or $\ell (\m_t)$. Using the conservation law, the subsystem size in Eq.~\eq{Relation:subsize} is rewritten as 
\begin{equation}
\fr{\ell (\mu_t)}{2 \m_t} =  \int_{\mu_t}^{\mu_{\L}} d\mu \frac{1}{\sqrt{\mu^2 - \mu_t^2}}  \left( \fr{1}{\m} \frac{dr}{d\mu} \right)  \, ,
\end{equation}
where $\m_\L = \m (r_\L) \to \infty$. This is the form of the Abel transformation (see \app{App: A}). Then,  $dr/d\m$ is determined by the inverse Abel transformation
\be
\frac{dr}{d\mu}  = - \fr{\m}{\pi} \fr{d}{d\m} \int_{\m}^{\m_\L} d \m_t \fr{\ell (\mu_t)}{\sqrt{\mu_t^2 - \m^2}}  .   \la{Relation: inverseAbel}
\ee

It was known that the entanglement entropy of a two-dimensional CFT has the following form
\be
S_E (\ell) = \fr{R}{2 G} \log \ell  + S_{\L} (\m_\L)  ,   \la{Input: entanglement}
\ee
where $S_{\L}  (\m_\L)$ means a UV divergent term independent of the subsystem size $\ell$. 
Now, we discuss how to reconstruct the dual geometry of this analytic entanglement entropy. Since the entanglement entropy is given as a function of the subsystem size, Eq.~\eq{Relation: ellmu} determines the turning point as a function of the subsystem size
\be
\m_t (\ell) = 4 G \fr{d S_E}{d \ell} = \fr{2 R}{\ell} . 
\ee
As a result, we finally determine the subsystem size as a function of the turning point
\be
\ell (\m_t) = \fr{2 R}{\m_t} .
\ee
Substituting this relation into the inverse Abel transformation in Eq.~\eq{Relation: inverseAbel} and evaluating it gives us the following result 
\be
\frac{dr}{d\mu}  = - \fr{2 R \m}{\pi} \fr{d}{d\m} \int_{\m}^{\m_\L} d \m_t \fr{1}{\m_t \sqrt{\mu_t^2 - \m^2}}  = \fr{R}{\m}  .
\ee
Solving this differential equation one more time, we finally obtain the metric factor of the dual geometry
\be
\m (r) = e^{r/R}  .
\ee
This is the metric of pure AdS space. A constant scalar profile is consistent with this solution and corresponds to a fixed point of the scalar dynamics. This example shows that the spatial dual geometry can be reconstructed analytically from the entanglement entropy in Eq.~\eq{Input: entanglement} within the stated metric ansatz. Appendix~\textcolor{blue}{B} gives a finite-temperature analytic example. When no closed form is available, the same inverse relation can be evaluated numerically.

\subsection{Relevant deformation generating a nontrivial RG flow}

We now consider a relevant deformation. The relevant deformation drives the UV CFT to a different IR fixed point. In the dual gravity description, a scalar field moves from an unstable extremum of the potential to a stable one and backreacts on the geometry \cite{deBoer:1999tgo, Verlinde:1999xm, deBoer:2000cz, Polchinski:2011im, Park:2021nyc}. We reconstruct the warp factor from the entanglement entropy and then use the bulk equations to determine the scalar profile and the sampled part of the potential. Fig.~\textcolor{blue}{2} (a) shows the entanglement entropy used as input. Fig.~\textcolor{blue}{2} (b) shows $dS_E/d\log\ell$. \tcr{Its UV and IR limits approach the logarithmic behavior expected at conformal fixed points. For the plot of the numerical derivative in Fig.~2(b), data with $\ell<10^{-7}$ are omitted to avoid the endpoint singularity. The Abel inversion below uses the full data set down to $\ell=\tcr{3.62430}\times10^{-14}$.}

%%%%%%%%%%%%%%%%%%%%%%%%
\begin{figure}[t]
\begin{center}
    % (a) Entanglement Entropy S(L)
    \subfigure[$S_E(\ell)$]{
        \includegraphics[height=5cm]{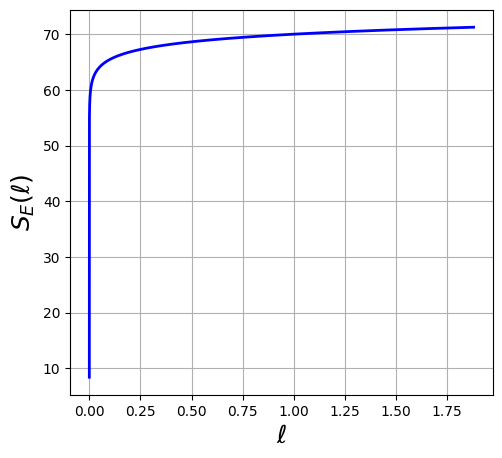} 
        \label{fig:input_S}
    }
    \hspace{0.3cm}
    % (b) Entropic C-function
    \subfigure[ $\fr{d S_E}{d \log \ell}$]{
        \includegraphics[height=5cm]{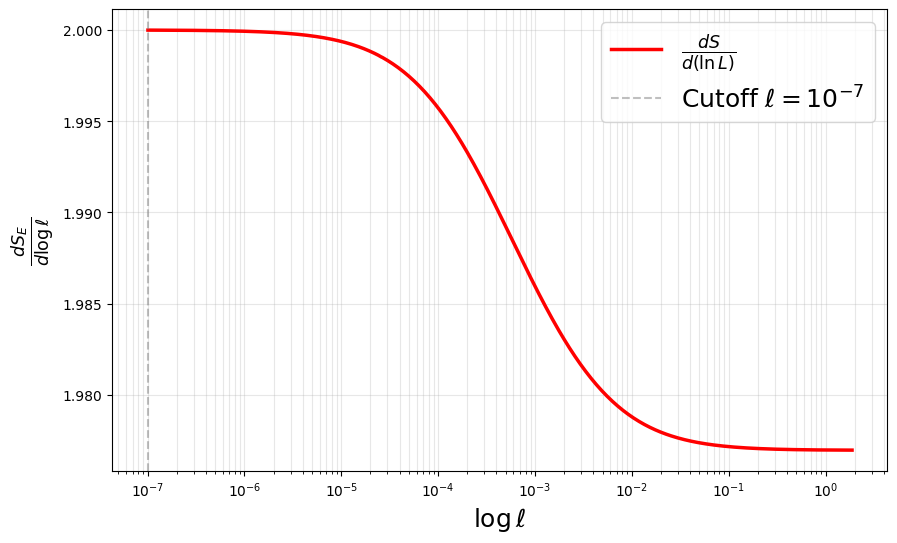} 
        \label{fig:input_C}
    }
    \caption{The input data for the reconstruction process. (a) The entanglement entropy $S_E(\ell)$ as a function of subsystem size $\ell$, generated from the theoretical model. (b) The entropic C-function $d S_E / d \ln \ell$. The monotonic decrease from the UV (small $\ell$) to the IR (large $\ell$) confirms the validity of the c-theorem and the existence of a proper RG flow. \tcr{The dashed line in (b) marks the lower end of the plotted range, $\ell=10^{-7}$. This cut is not used in the reconstruction.}}
    \label{fig:input_data}
\end{center}
\end{figure}
%%%%%%%%%%%%%%%%%%%%%%%%%

Applying the inverse Abel transformation in Eq.~\eq{Relation: inverseAbel},  the radial coordinate is determined as a function of the metric factor 
\be
r (\m_\L)  - r (\m)  &=& - \fr{1}{\pi} \int_{\m}^{\m_\L} d \td{\m} \,  \td{\m} \   \fr{d}{d \td{\m}} \int_{\td{\m}}^{\m_\L} d \m_t \fr{\ell (\mu_t)}{\sqrt{\mu_t^2 - \td{\m}^2}}   \nn
&=& \frac{1}{\pi} \int_{\mu}^{\mu_{\L}} d\mu_t \ \ell(\mu_t) \left[ \frac{\mu}{ \sqrt{\mu_t^2 - \mu^2}} + \arccos\left( \frac{\mu}{\mu_t} \right) \rb \,,  \la{Formula: radialco}
\ee
where we use partial integration to obtain the last relation. Once $\ell(\m_t)$ is known, Eq.~\eq{Formula: radialco} gives $r(\m)$. We obtain $\m_t(\ell)$ from $S_E(\ell)$ by Eq.~\eq{Relation: ellmu}, invert it numerically, and evaluate the Abel integral. This gives $A(r)=\log\m(r)$. Fig.~\textcolor{blue}{3} shows $A(r)$ and $dA/dr$.

%\subsection{Reconstruction of $\phi(r)$ and $V(\phi)$}

With the reconstructed metric factor $A(r)$, we can further determine the scalar field profile and the scalar potential. The dynamics of the dual gravity are governed by the equations of motion derived from Eq.~\eq{Ansatz: dual gravity}
\be
0 &=& 4 \dot{A}^2 - \dot{\phi}^2 - \fr{4}{R^2} + \fr{2 V(\ph)}{R^2} , \la{eq:eom_1} \\
0 &=& 4 \ddot{A} + 4 \dot{A}^2 + \dot{\phi}^2 - \fr{4}{R^2} + \fr{2 V(\ph)}{R^2} , \la{eq:eom_2} \\
0 &=& \ddot{\phi} + 2 \dot{A} \dot{\phi} - \fr{1}{R^2} \fr{\pa V(\ph)}{\pa \ph} . \la{Equation:Einstein}
\ee
where the dot means a derivative with respect to the radial coordinate $r$. Here, the second and third equations govern the dynamics of $A$ and $\ph$, while the first is a constraint of the gravity theory. Since they are not independent, the gravity solution can be determined by only two of them. Subtracting Eq.~\eq{eq:eom_1} from Eq.~\eq{eq:eom_2} yields a direct relation between the metric and the scalar field kinetic term.
\begin{equation}
   \dot{\phi} = \pm \sqrt{-2\ddot{A}} \,.
\end{equation}
This relation implies that the null energy condition requires $\ddot{A}\le0$. For a smooth physical reconstruction we therefore integrate
\begin{equation}
   \ph(r)=\ph_\L\pm\sqrt{2}\int_r^{r_\L}d\td r\,\sqrt{-\ddot A(\td r)}\,.
\end{equation}
In Fig.~\textcolor{blue}{4} (a), we depict the scalar field profile $\ph(r)$ derived from the reconstructed $A(r)$.
Once $\phi(r)$ and $A(r)$ are known, Eq.~\eq{eq:eom_1} determines the scalar potential $V(r)$ as a function of the radial coordinate
\begin{equation}
    V(r)=2-2\,R^2\,\dot{A}^{\,2}-R^2\,\ddot{A}.  
\end{equation}
Inverting $\phi(r)$ gives $r(\phi)$ and therefore the scalar potential $V(\phi)$ shown in Fig.~\textcolor{blue}{4} (b).

%%%%%%%%%%%%%%%%%%%%%%%%%
\begin{figure}[t]
\begin{center}
    \subfigure[Reconstructed $A(r)$]{
        \includegraphics[height=4.7cm]{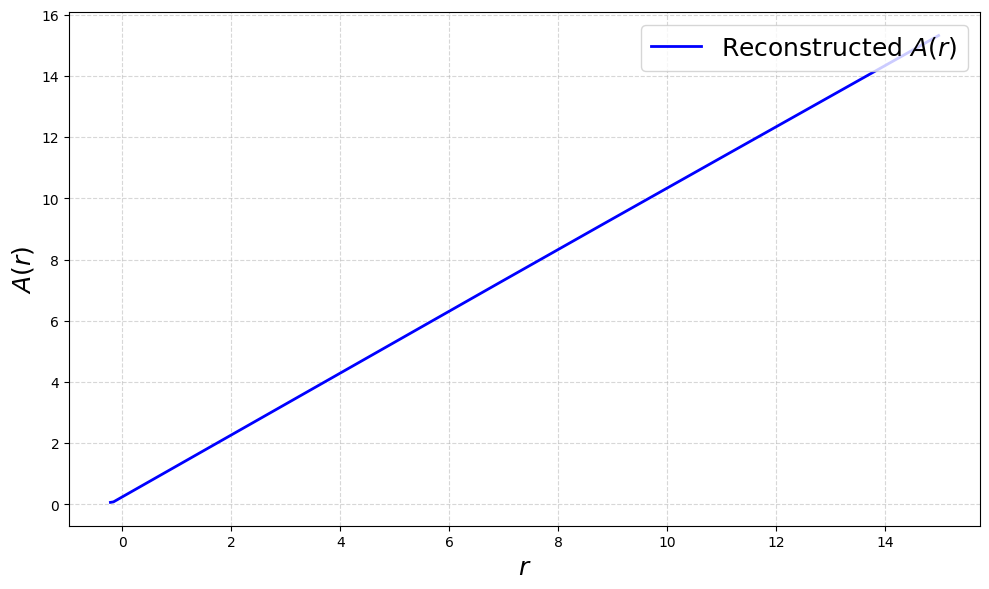} 
        \label{fig:recon_A}
    }
    \hspace{0.0cm}
    \subfigure[$\dot{A} = \frac{dA(r)}{dr} $]{
        \includegraphics[height=4.7cm]{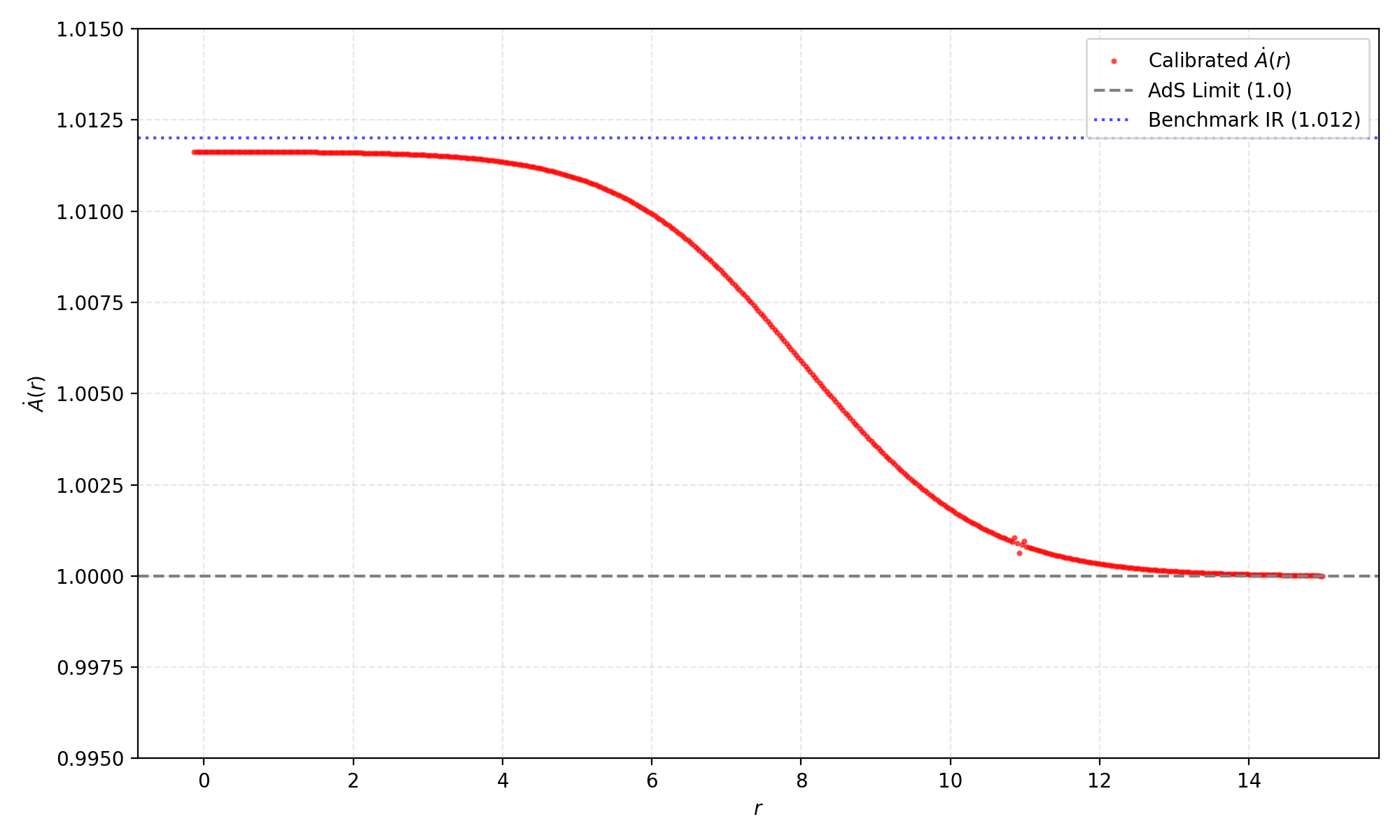} 
        \label{fig:recon_A'}
    }
    \caption{The numerical reconstruction results of the metric. (a) The reconstructed warp factor $A(r)$ in the range $r < 15$. (b) The derivative $\dot{A} = d A/dr$, showing the stability of the reconstruction near the AdS limit ($\dot{A} \approx 1$).}
    \label{fig:reconstruction_result_A}
\end{center}
\end{figure}
%%%%%%%%%%%%%%%%%%%%%%%%%

To obtain the entanglement entropy data in Fig.~\textcolor{blue}{2} (a), we use a specific scalar potential form
\be
V (\ph)  =     \fr{m_\ph^2}{2} \ph^2 + \fr{\l}{4} \ph^4 ,      \la{Result:scalarpot}
\ee
where $m_\ph^2$ and $\l$ are dimensionless. According to the AdS/CFT dictionary, this bulk scalar mass is related to the conformal dimension $\Delta$ of the deformation operator \cite{Gubser:1998bc, Witten:1998qj, Witten:1998zw}
\begin{equation}
    m_\ph^2 = \Delta (\Delta - 2) \,.  \la{Relation: condim}
\end{equation}
When $m_\ph^2>0$ for $\l>0$, the deformation is irrelevant and the scalar potential has a stable minimum at the UV fixed point. For $m_\ph^2<0$, the deformation is relevant and the potential has a local maximum at $\ph=0$. The nonzero extrema satisfy $\ph_\pm=\pm\sqrt{-m_\ph^2/\l}$. We consider the branch $\ph>0$. For the data in Fig.~\textcolor{blue}{2} (a), we use $m_\ph^2=-3/4$, $\l=3$, $R=1$, and $G=1/4$. \tcr{In the reconstruction procedure, we parametrize the reconstructed scalar potential along the RG trajectory as
\begin{equation}
V(\phi)=c_0+c_1\phi+c_2\phi^2+c_3\phi^3+c_4\phi^4 \, .
\label{Result:reconstructed_potential_fit}
\end{equation}
The benchmark AdS vacuum satisfies $V(0)=0$, but we do not impose this condition on the fit. An unconstrained fit of the five coefficients to the 481 reconstructed points gives $(c_0,c_1,c_2,c_3,c_4)=(-0.00003,\allowbreak -0.00790,\allowbreak -0.37224,\allowbreak 0.03761,\allowbreak 0.72944)$, which is compared with the benchmark coefficients $(0,\allowbreak 0,\allowbreak -0.375,\allowbreak 0,\allowbreak 0.750)$ in Eq.~\eq{Result:scalarpot}. The small nonzero values of \(c_0\), \(c_1\), and \(c_3\) are interpreted as numerical errors in the reconstruction procedure. The ultraviolet and noise sensitivities of the reconstruction are summarized in Appendix~\ref{AppC}.}

%%%%%%%%%%%%%%%%%%%%%%%%%
\begin{figure}[t]
\begin{center}   
 \hspace{-0.5cm}
    \subfigure[$\phi(r)$]{
        \includegraphics[height=4.5cm]{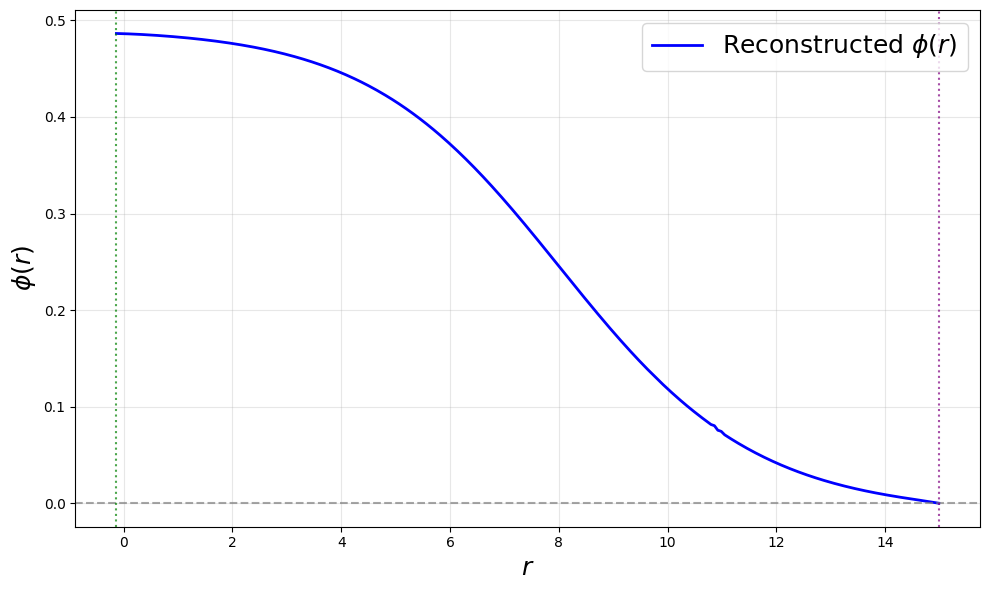} 
        \label{fig:recon_phi}
    }
    \hspace{-0.5cm}
    % (b) Reconstructed Potential V(phi)
    \subfigure[$V(\phi)$]{
        \includegraphics[height=4.7cm]{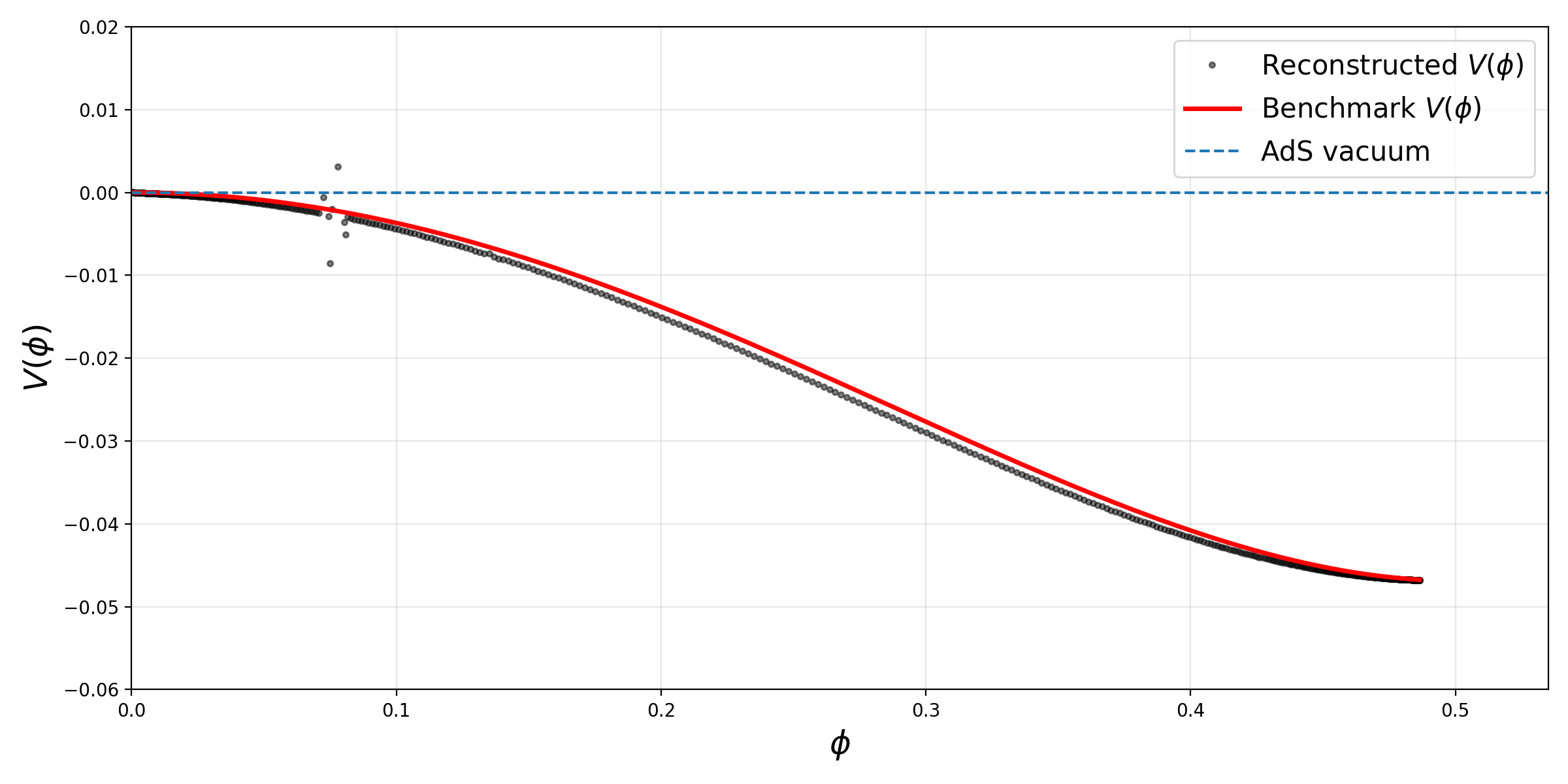} 
        \label{fig:recon_V}
    }
    \caption{Reconstruction of the scalar sector. (a) The scalar profile $\phi(r)$. (b) The reconstructed scalar potential $V(\phi)$ is shown by the numerical points and the benchmark potential of Eq.~\eq{Result:scalarpot} by the red line. The dashed line denotes the AdS vacuum value $V=0$. The reconstructed potential is parametrized by Eq.~\eq{Result:reconstructed_potential_fit}.}
    \label{fig:reconstruction_result}
\end{center}
\end{figure}
%%%%%%%%%%%%%%%%%%%%%%%%%

%%%%%%%%%%%%%%%%%%%%%%%%%
\begin{figure}[t]
\begin{center}
    \subfigure[$\beta$-function]{
        \includegraphics[height=4.7cm]{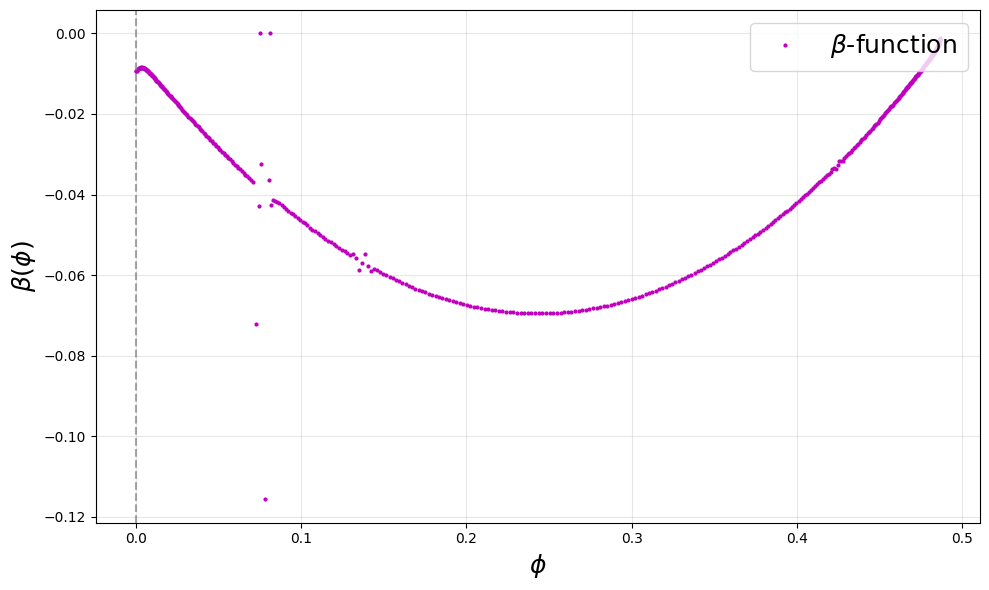} 
        \label{fig:beta_function}
    }
    \hspace{0.0cm}
    % (b) Reconstructed Potential V(phi)
    \subfigure[$c$-function]{
        \includegraphics[height=4.7cm]{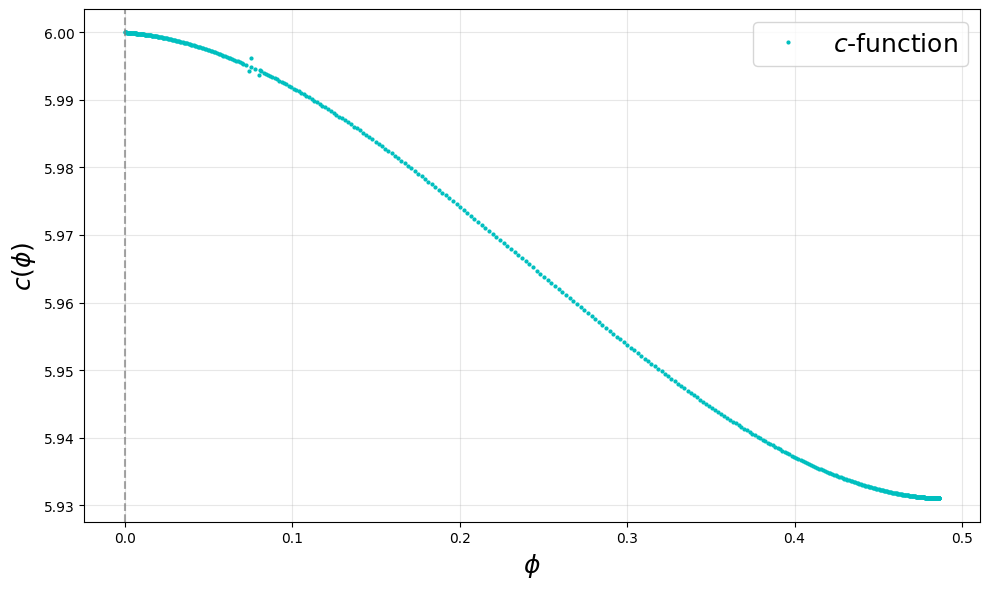} 
        \label{fig:c_function}
    }
    \caption{(a) The $\beta$-function as a function of the coupling constant $\ph$. (b) The holographic $c$-function as a function of the coupling constant $\ph$.}
\end{center}
\end{figure}

%%%%%%%%%%%%%%%%%%%%%%%%%

The reconstructed coefficient gives $m_\ph^2=2c_2=\tcr{-0.74448}$. To understand this result, we note that Eq.~\eq{Relation: condim} has two solutions, $\Delta_\pm=1\pm\sqrt{1+m_\ph^2}$, with $\Delta_+\simeq\tcr{1.50549}$ and $\Delta_-\simeq\tcr{0.49451}$. Near the boundary, $\ph(z)=\ph_-z^{\Delta_-}+\ph_+z^{\Delta_+}+\cdots$. We use standard quantization, for which $\ph_-$ is the source, $\ph_+$ is the response, and the operator dimension is $\Delta=\Delta_+>1$. Alternative quantization assigns $\Delta_-$. Numerically we impose $\ph(r_\L)=0$ and choose the positive monotonic branch that flows from $\ph=0$ to the exact IR fixed point $\ph_{\rm IR}=1/2$. These choices are model and boundary inputs, and the source amplitude $\ph_-$ is not independently inferred from entanglement data. The standard-quantization result agrees with the input value $\Delta=3/2$ within the numerical error. Within this assumed holographic dictionary, the metric factor $\m$ and scalar mode define the RG scale and running coupling, so we obtain the $\beta$-function \cite{Park:2021nyc}
\be
\b (\ph) = \fr{d \ph(\m)}{d \log \m}  .
\ee
Fig.~\textcolor{blue}{5} (a) shows the resulting $\beta$-function. It vanishes at the UV and IR fixed points, $\ph=0$ and $\ph=1/2$, respectively. Between the two points it remains negative for the positive branch of the relevant deformation. The holographic $c$-function is defined by \cite{Zamolodchikov:1986gt, Gubser:2000nd, Deger:2002hv}
\be
c = \fr{3}{2G} \fr{1}{\dot{A}}  .
\ee
Fig.~\textcolor{blue}{5} (b) shows that the holographic $c$-function decreases monotonically from $\ph=0$ to $\ph=0.5$, in agreement with the $c$-theorem.

%%%%%%%%%%%%%%%%%%%%%%%%%%% 
%                                                                                 %
%                                Sec.  Discussion                      %
%                                                                                 %
%%%%%%%%%%%%%%%%%%%%%%%%%%% 

\section{Discussion}

We study an inverse holographic problem using entanglement entropy as the boundary input. At finite temperature, equal-time entanglement entropy determines the spatial radial geometry within the assumed metric ansatz. The temporal component is not fixed by this data alone. Therefore, the Hawking temperature and the thermodynamic quantities derived from it require an additional temporal bulk completion. Once this completion and the matter model are specified, we show that the reconstructed geometry gives thermodynamic quantities consistent with the assumed thermal system.

For a zero-temperature relevant deformation, the Abel inversion directly determines the warp factor from the entanglement entropy. The Abel inversion itself follows the inverse structure developed in previous studies. In the Einstein-scalar model considered here, we then use the bulk equations to determine a monotonic scalar profile and the part of $V(\phi)$ traced by the RG flow. In this sense, the reconstruction goes beyond the spatial geometry and gives information about the scalar sector. However, this does not imply that entanglement entropy selects a unique gravitational action. The result is a reconstruction inside the stated Einstein-scalar model class and only the part of the potential sampled by the flow is determined. After finding a solution satisfying the equations of motion of the assumed gravity theory, we show that the reconstruction reproduces the expected scalar potential with small errors and gives the $\beta$-function and $c$-function describing the flow between the UV and IR fixed points. \tcr{This conclusion applies to smooth theoretical data with sufficiently deep UV coverage, because the present procedure is highly sensitive to noise (Appendix~\ref{AppC}).}

These results show that, within the assumed bulk model and holographic dictionary, entanglement entropy can provide information beyond the reconstructed spatial metric. Extensions to higher-dimensional systems, defect CFTs \cite{Park:2024pkt, Linardopoulos:2026mut}, and finite-temperature RG flows with scalar hair are interesting future directions. More general geometries may require additional boundary data.

\vspace{\baselineskip}

{\bf Acknowledgement}

This work was supported by the National Research Foundation of Korea (NRF) grant funded by the Korea government (MSIT) (No. RS-2026-25473667). 

\vspace{0.1cm}

%\noindent{\large \bf Appendix}
\appendix
\section{Abel transformation \la{App: A}}

Assume that a function $F(y) $ is expressed as an integral of an unknown function $f(r)$
\be
F (y) = \int_{y}^{\infty} d r \fr{f(r)}{\sqrt{r^2 - y^2}} ,   \la{Relation: Abel}
\ee
which is known as the Abel transformation. Then, the unknown function $f(r)$ can be determined by the inverse Abel transformation \cite{Gorenflo1991}
\be
f (r) =  - \fr{2 r}{\pi}  \int_r^{\infty } d y \ \fr{1}{\sqrt{y^2 - r^2}} \  \fr{d F (y)}{dy} 
= - \fr{2}{\pi} \fr{d}{d r} \int_r^{\infty } d y \ \fr{y \, F(y)}{\sqrt{y^2 - r^2}}   ,   \la{Relation: InAbel}
\ee
where we assume
\be
\lim_{y \to \infty} \ y \, F(y) = 0 .   \la{Condition: uniqueness}
\ee
This relation indicates that the inverse Abel transformation in Eq.~\eq{Relation: InAbel} is the solution of the integral equation in Eq.~\eq{Relation: Abel}.

To prove the Abel transformation, we substitute Eq.~\eq{Relation: InAbel} into Eq.~\eq{Relation: Abel} and calculate it
\be
{\rm \eq{Relation: Abel}}  &=& - \fr{2}{\pi} \int_{y}^{\infty} d r  \fr{r}{\sqrt{r^2 - y^2}} \int_r^{\infty } d \bar{y} \ \fr{1}{ \sqrt{\bar{y}^2 - r^2}}  \fr{d F (\bar{y})}{d \bar{y}} \nn
 &=& - \fr{2}{\pi} \int_y^{\infty } d \bar{y} \  \fr{d F (\bar{y})}{d \bar{y}}
 \int_{y}^{\bar{y}} d r  \fr{r}{\sqrt{r^2 - y^2} \sqrt{\bar{y}^2 - r^2} } \nn
&=& -  \int_y^{\infty } d \bar{y} \  \fr{d F (\bar{y})}{d \bar{y}}  = F (y)  .
\ee
This result proves that Eq.~\eq{Relation: InAbel} automatically satisfies the Abel transformation in Eq.~\eq{Relation: Abel}. The Abel transformation is useful for deriving the dual geometry from entanglement entropy without resorting to heavy numerical simulations. 

When $F(y)$ is continuously differentiable and satisfies Eq.~\eq{Condition: uniqueness}, the Abel inversion is mathematically unique. The physical reconstruction requires two further conditions. We require $\mu_t(\ell)$ to be monotonic so that the inverse function $\ell(\mu_t)$ is single valued. We also require a monotonic scalar profile so that $\phi(r)$ can be inverted and the potential can be written as $V(\phi)$. These are conditions of the reconstruction rather than information supplied by entanglement entropy alone. Under these conditions the Abel inversion and the subsequent reconstruction are unique only within the stated Einstein-scalar ansatz and over the scalar field interval sampled by the flow. We therefore do not claim uniqueness of the underlying gravitational action from entanglement entropy alone. Different bulk actions may support the same reconstructed background geometry.

%\subsection{Holographic dual of CFT at zero temperature}
\section{Inverse Abel transformation of the entanglement entropy \la{App: B}}

To apply the Abel transformation, we consider a static and translationally invariant three-dimensional geometry in the radial gauge
\be
ds^2 =  \fr{r^2}{R^2}  \ls - \ F (r) dt^2  + dx^2 \rs + \fr{R^2}{r^2} \fr{1}{ G(r)} dr^2 ,
\ee
where $F(r)$ and $G(r)$ are undetermined functions. Assuming the holographic relation, the entanglement entropy of the dual QFT is governed by
\be
S_E = \fr{1}{4 G} \int_{-\ell/2}^{\ell/2}  dx \sqrt{\fr{r^2}{R^2} + \fr{R^2}{r^2 G(r)} \ls \fr{dr}{dx} \rs^2} .
\ee
For an equal-time interval, this functional depends on the spatial radial metric function $G(r)$ but not on the temporal function $F(r)$. Therefore the Abel inversion below reconstructs $G(r)$ only. The function $F(r)$ remains undetermined by equal-time entanglement entropy. Relating $F(r)$ to $G(r)$ requires additional bulk equations/model assumptions or a Schwarzschild-gauge choice.
Using the translational invariance in the $x$-direction, the subsystem size and the entanglement entropy are reexpressed as the radial integrals
\be
\fr{\ell (r_t)}{2 R^2 r_t } &=& \int_{r_t}^{\infty} dr \fr{1}{r^2 \sqrt{G(r)} \sqrt{r^2 -r _t^2}} , \la{Relation: subsell} \\
S_E (r_t) &=& \fr{R}{2 G} \int_{r_t}^{\infty} dr \fr{1}{ \sqrt{G(r)} \sqrt{r^2 -r _t^2}} , \la{Relation: subsE}
\ee 
where $r_t$ indicates a turning point satisfying $d r/d x = 0$ at $r=r_t$. 

Recalling that the integrals in Eq.~\eq{Relation: subsell} and Eq.~\eq{Relation: subsE} are in the form of the Abel transformation, the undetermined metric factor $G(r)$ can be fixed by the inverse Abel transformation
\be
\fr{1}{\sqrt{G(r)}} = - \fr{r^2}{\pi R^2} \fr{d}{dr} \int_{r}^{\infty} d r_t \ \fr{\ell(r_t)}{\sqrt{r_t^2 -r^2}} .  \la{Equation: inverseAbelt}
\ee 
To determine $G(r)$ by this procedure, we need $\ell(r_t)$. If the entanglement entropy $S_E(\ell)$ is known, this function follows from the derivative of the entropy. To see this, we take into account alternative expressions for the subsystem size and the entanglement entropy 
\be
 \ell (r_t)  &=& \int_{- \ell/2}^{\ell/2}  dx   = 2 \int_{r_t}^{\infty} dr \ \fr{dx}{dr}   , \\
S_E (r_t) &=& \fr{1}{2 G} \int_{r_t}^{\infty} dr \sqrt{\fr{r^2}{R^2}  \ls \fr{dx}{dr} \rs^2 + \fr{R^2}{r^2 G(r)}} .
\ee 
Recalling that $dx/dr \to \infty$ at the turning point, these alternative relations lead to the following identity relation
\be
\fr{d S_E}{d \ell} = \fr{d S_E/ d r_t}{d \ell / d r_t }  =\fr{1}{4 G} \fr{r_t (\ell)}{R} .   \la{Equation: subsizert}
\ee 
From this result, the size of the subsystem can be expressed as a function of the turning point $\ell(r_t)$, which in turn allows us to determine the metric function $G(r)$ in Eq.~\eq{Equation: inverseAbelt}.

\subsection{For CFT at zero temperature}

At the critical point with the conformal symmetry, the entanglement entropy has the following analytic form \cite{Ryu:2006bv, Ryu:2006ef, Park:2015dia}
\be
S_E = \fr{R}{2 G} \log \ls \fr{\ell r_\L}{R^2} \rs ,
\ee
where $r_\L \to \infty$ is an appropriate UV cutoff. Applying Eq.~\eq{Equation: subsizert}, we find the subsystem size as a function of the turning point
\be
\ell(r_t) = \fr{2 R^2}{ r_t}  .
\ee
Plugging this into Eq.~\eq{Equation: inverseAbelt} 
\be
\fr{1}{\sqrt{G(r)}} = - \fr{2 r^2}{\pi } \fr{d}{dr} \int_{r}^{\infty} d r_t \ \fr{1}{r_t \sqrt{r_t^2 -r^2}} , 
\ee
and performing the integration, we finally obtain the following result
\be
G(r) =1 .
\ee
This is the spatial radial metric function of AdS. Thus the critical entanglement entropy reconstructs $G(r)=1$. Identifying the full spacetime as AdS additionally uses the temporal completion $F(r)=1$ within the assumed Lorentz-invariant/Einstein-gravity setting. 

\subsection{For thermal CFT}

We now take into account a more nontrivial example. For a two-dimensional CFT at finite temperature, the entanglement entropy is given by \cite{Ryu:2006bv, Ryu:2006ef, Kim:2016jwu}
\be
S_E = \fr{c}{3} \log \lb \fr{  r_\L}{\pi R^2 T_H} \sinh \ls \pi T_H \ell \rs \rb ,
\ee
where $r_\L$ is an appropriate UV cutoff and $R$ is introduced for the correct dimension counting. In the IR limit ($\ell \to \infty$), the entanglement entropy is reduced to
\be
S_E =  \fr{\ell}{2 G}  \pi R T_H ,
\ee
where we use the holographic relation, $c = 3 R/(2 G)$ and $V \sim \ell$ in the IR limit. Recalling that the thermal entanglement entropy approaches a thermal entropy \cite{Kim:2016jwu} and comparing this with the Bekenstein-Hawking entropy in Eq.~\eq{Relation: BHentr} for $d=2$, we determine the temperature as a function of a black hole horizon 
\be
T_H = \fr{r_h}{2 \pi R^2} ,
\ee
Here $T_H$ is the thermal parameter appearing in the boundary CFT entropy. Its interpretation as the Hawking temperature of a full bulk spacetime requires the additional temporal-metric completion discussed below. Using this relation, we can reexpress the above entanglement entropy in terms of the dual gravity theory's variables
\be
S_E = \fr{R}{2 G} \log \lb \fr{2 r_\L}{r_h} \sinh \ls \fr{r_h \ell }{2 R^2}\rs \rb .  \la{Result: fintemEE}
\ee
At this stage, conformal symmetry suggests a BTZ completion \cite{Banados:1992wn}.

Applying Eq.~\eq{Equation: subsizert}, the subsystem size is determined in the following form
\be
\ell (r_t) = \fr{R^2}{r_h} \log \ls \fr{r_t + r_h}{r_t - r_h}\rs .
\ee 
Plugging this into the inverse Abel transformation, the metric factor can be written as
\be
\fr{1}{\sqrt{G(r)}} = - \fr{r^2}{\pi r_h} \fr{d \, I (r)}{dr}  ,
\ee
where $I$ is defined as
\be
I (r) = \int_{r}^{\infty} d r_t \ \fr{1}{\sqrt{r_t^2 -r^2}} \  \log \ls \fr{r_t + r_h}{r_t - r_h}\rs .
\ee
Performing this integral with successive changes of variable, we finally arrive at
\be
I (r) = 2 \pi \lb \arctan \tcr{ \sqrt{\fr{r+r_h}{r-r_h}}} - \fr{\pi}{4} \rb = \pi \arcsin \ls \fr{r_h}{r} \rs . 
\ee
Finally, we derive the following spatial radial metric function from the thermal entanglement entropy in Eq.~\eq{Result: fintemEE}
\be
G(r)  =  1 - \fr{r_h^2}{r^2} .
\ee
This is precisely the radial metric function of the BTZ geometry. If we impose the Schwarzschild metric ansatz, we obtain $F(r)=G(r)$, which leads to the Schwarzschild/BTZ completion within the assumed Einstein-gravity model. As a result, this analytic example shows that the entanglement entropy data can reconstruct the dual gravity within the appropriate assumptions.

%\vspace{0.5cm}

%%%%%%%%%%%%%%%%%%%%%%%%%%%
%                                                                                  %
%                          The Bibliography                           %
%                                                                                  %
%%%%%%%%%%%%%%%%%%%%%%%%%%%

%\bibliographystyle{apsrev4-1}
%\bibliography{References}


\tcr{ \section{Scalar-potential reconstruction and sensitivity tests} \la{AppC}}

\tcr{In the convention of Eq.~\eq{Ansatz: dual gravity}, the benchmark potential becomes $V(\phi)=-0.375\phi^2+0.750\phi^4$. We reconstruct this benchmark potential with the five coefficients $c_0,\ldots,c_4$ in Eq.~\eq{Result:reconstructed_potential_fit}. Before the polynomial fit, we discard points whose numerically evaluated $\dot A$ lies outside $0.99<\dot A<1.02$. We refer to this selection as the $\dot A$ window filter. We then normalize $\dot A$ by its mean over $14.5\leq r\leq15.5$ and use the resulting data at $r\leq15$ for the scalar-field and potential reconstruction.}

\tcr{Using the resulting 481 reconstructed points gives $(c_0,c_1,c_2,c_3,c_4)=(-0.00003,\allowbreak -0.00790,\allowbreak -0.37224,\allowbreak 0.03761,\allowbreak 0.72944)$ with the coefficient of determination $R^2_{\rm fit}=0.99958$ defined as
\begin{equation}
R^2_{\rm fit}
=
1-\frac{\sum_i\big(V_i-V(\phi_i)\big)^2}
        {\sum_i\big(V_i-\bar V\big)^2}.
\end{equation}
Here, $V_i$ are the reconstructed values of the potential, $V(\phi_i)$ is the fitted polynomial in Eq.~\eq{Result:reconstructed_potential_fit}, and $\bar V$ is the mean of $V_i$. Thus $R^2_{\rm fit}$ measures how well the polynomial represents the reconstructed points, not how close they are to the benchmark potential. An additional local numerical filter leaves 472 points and changes the result only slightly to $(-0.00003,\allowbreak -0.00792,\allowbreak -0.37202,\allowbreak 0.03695,\allowbreak 0.73005)$. The quadratic and quartic coefficients remain close to the benchmark values $c_2=-3/8$ and $c_4=3/4$.}

\tcr{Table~2 shows the dependence of the reconstructed coefficients on the lower entanglement cutoff $\ell_{\min}$, below which the input data are discarded. The same Abel inversion, $\dot A$ window filter, scalar reconstruction, and quartic parametrization are used for every row, while the local filter leaving 472 points is not applied in Tables~2 and 3. Here $E_V$ is the RMS deviation from the benchmark potential over $0.10\leq\phi\leq0.45$. The reconstructed coefficients remain nearly unchanged for $\ell_{\min}\leq10^{-9}$ and change only modestly at $10^{-8}$, whereas the reconstructions with $\ell_{\min}=10^{-7}$ and $10^{-6}$ no longer reproduce the benchmark potential faithfully. The large $R_{\rm fit}^2$ at $10^{-7}$ ($R_{\rm fit}^2=0.99798$) also shows that $R_{\rm fit}^2$ alone is not a sufficient criterion for recovering the bulk potential.}

\vspace{0.5cm}
\begin{center}
\small
\begin{tabular}{|c||c|c|c|c|c|c|}
\hline
$\ell_{\min}$ & $c_0$ & $c_1$ & $c_2$ & $c_3$ & $c_4$ & $E_V$ \\
\hline\hline
$3.62430\times10^{-14}$ & $-0.00003$ & $-0.00790$ & $-0.37224$ & $0.03761$ & $0.72944$ & $0.00111$ \\
\hline
$10^{-12}$ & $-0.00003$ & $-0.00790$ & $-0.37224$ & $0.03761$ & $0.72943$ & $0.00111$ \\
\hline
$10^{-10}$ & $-0.00003$ & $-0.00790$ & $-0.37222$ & $0.03754$ & $0.72950$ & $0.00111$ \\
\hline
$10^{-9}$ & $-0.00003$ & $-0.00790$ & $-0.37227$ & $0.03769$ & $0.72936$ & $0.00111$ \\
\hline
$10^{-8}$ & $-0.00004$ & $-0.00778$ & $-0.37155$ & $0.03481$ & $0.73121$ & $0.00109$ \\
\hline
$10^{-7}$ & $-0.00196$ & $-0.17158$ & $0.83340$ & $-2.83823$ & $2.81767$ & $0.00680$ \\
\hline
$10^{-6}$ & $0.02442$ & $-0.79821$ & $3.63562$ & $-7.82818$ & $5.83027$ & $0.02613$ \\
\hline
\end{tabular}
\end{center}
\vspace{0.3cm}
\noindent {{\bf Table 2}. UV-cutoff dependence of the reconstructed coefficients.}\label{tabuvscan}
\par\vspace{0.45cm}

\tcr{We test the sensitivity of the inverse reconstruction to controlled Gaussian perturbations of the boundary entanglement data,
\begin{equation}
S_E^{\rm noisy}(\ell_i)
=
S_E(\ell_i)+\sigma\Delta S_E\,\xi_i,
\qquad
\Delta S_E
=
\max_{\ell_i\geq10^{-7}} S_E(\ell_i)
-
\min_{\ell_i\geq10^{-7}} S_E(\ell_i).
\end{equation}
Here, the $\xi_i$ are independent standard normal random variables, and $\Delta S_E$ is the range of the unperturbed entropy values over the sampled points to which noise is added. Thus, the additive noise at each perturbed point has standard deviation $\sigma\Delta S_E$. Noise is added to the 530 sampled points with $1.01034\times10^{-7}\leq\ell\leq1.87946$, for which $\Delta S_E=33.29145$. The remaining 470 points with $\ell<10^{-7}$ are kept unperturbed and retained in the inversion. Consequently, the $\sigma=0$ row of Table~3 coincides with the full-data reference reconstruction. For each nonzero noise level, we perform 20 independent realizations using the same reconstruction procedure and repeat the unconstrained five-coefficient fit. Table~3 reports the median values of $c_3$ and $R_{\mathrm{fit}}^2$ across these realizations. At $\sigma=10^{-8}$, the median value of $c_3$ already departs appreciably from its unperturbed value, and the median $R_{\mathrm{fit}}^2$ decreases, even though the deep-UV data remain unperturbed. At larger noise levels, the median $R_{\mathrm{fit}}^2$ falls further, indicating that the quartic polynomial no longer provides a good description of the reconstructed points. The smaller median values of $c_3$ at these noise levels therefore do not indicate improved agreement with the benchmark potential. These tests show that the present reconstruction procedure is highly sensitive to noise in the entanglement data. We accordingly restrict our conclusions to reconstruction from smooth theoretical entanglement data with sufficiently deep UV coverage. Extending the method to data with appreciable noise would require additional regularization and a separate assessment of its accuracy and stability.}

\vspace{0.5cm}
\begin{center}
\begin{tabular}{|c||c|c|}
\hline
$\sigma$ & median $c_3$ & median $R_{\rm fit}^2$ \\
\hline\hline
$0$ & $0.03761$ & $0.99958$ \\
\hline
$10^{-8}$ & $0.49617$ & $0.82580$ \\
\hline
$10^{-7}$ & $0.01619$ & $0.04603$ \\
\hline
$10^{-6}$ & $-0.02708$ & $0.02146$ \\
\hline
\end{tabular}
\end{center}
\vspace{0.3cm}
\noindent {{\bf Table 3}. Sensitivity of the reconstructed coefficients to entanglement perturbations. }\label{tabnoisescan}
\par%

\end{document}